\documentclass[a4paper,fleqn]{cas-dc}

\usepackage[numbers]{natbib}
\usepackage{mathtools}
\usepackage{amsmath}
\usepackage{pdflscape}
\def\tsc#1{\csdef{#1}{\textsc{\lowercase{#1}}\xspace}}
\tsc{WGM}
\tsc{QE}
\newcommand{\ud}{\mathrm{d}}

\begin{document}
\let\WriteBookmarks\relax
\def\floatpagepagefraction{1}
\def\textpagefraction{.001}

\shorttitle{Information entropy and fragmentation functions}    

\shortauthors{Benito-Calvi\~no, Garc\'{\i}a-Olivares and Llanes-Estrada}  

\title [mode = title]{Information entropy and fragmentation functions}

\author[1]{Guillermo Benito-Calvi\~no}
\author[1]{Javier Garc\'{\i}a-Olivares}
\author[1]{Felipe J. Llanes-Estrada}


\affiliation[1]{organization={Univ. Complutense de Madrid},
            addressline={Fac. Ciencias F\'{\i}sicas, Dept. F\'{\i}sica Te\'orica e IPARCOS, Plaza de las Ciencias 1}, 
            city={Madrid},
            postcode={28040}, 
            country={Spain}}

\begin{abstract}
Several groups have recently investigated the flow of information in high-energy collisions, 
from the entanglement entropy of the proton yielding classical Shannon entropy of its parton distribution functions (pdfs), through jet splitting generating entropy, to the entropy distribution in hadron decays. \\
Lacking in the literature is a discussion of the information entropy of fragmentation functions (FFs) in the instances where they can be considered as probability distributions, and we here provide it. We find that this entropy is 
a single, convenient number to characterize future progress in the extraction of fragmentation functions. \\
We also deploy the related Kullback-Leibler divergence between two distributions to assess existing relations among FFs and parton distribution functions (pdfs) such as that of Barone, Drago and Ma. From a couple of current parametrizations of FFs, we do not find supporting empirical evidence for the relation, although it is possible that FFs and pdfs have similar power-laws near the $x=1$ endpoint.
\end{abstract}

\maketitle

\section{Introduction}\label{sec:intro}
\subsection{Information entropy in particle physics} \label{subsec:info}

The rapid expansion of quantum information theory~\cite{Galindo:2001ei} has made groups in high-energy physics aware of 
its possible uses to think about particle accelerator data. Several applications include the distinction among conventional hadrons and exotica~\cite{Colangelo:2018mrt}, the exploration of supersymmetric models~\cite{Gupta:2020whs}, 
predictions on exotic decays of the Higgs boson~\cite{Alves:2020cmr} based on the observation that the Higgs-boson mass at $125$ GeV plays a special role by maximizing the product of decay probabilities~\cite{dEnterria:2012eip}, which lead to the formulation of a maximum-entropy hypothesis for the Higgs boson~\cite{Alves:2014ksa} 
or the analysis of heavy-ion collisions~\cite{Ma:2018wtw}.

A case of particular interest is information entropy in high-energy collisions. The initial states of collider experiments are usually prepared in pure quantum states where a few particles (for example, an electron and a proton or two hadrons) have well defined momenta. However, when they collide, cross-sections can often be described in collinear factorization in terms of parton distribution functions that have a classical probability distribution. The conversion of the entanglement entropy in the initial state wavefunction to the final pdf classical information entropy has been authoritatively described a short time ago~\cite{Kharzeev:2017qzs}, and studies continue~\cite{Kou:2022dkw}. If a jet forms in such collisions, entropy evolves~\cite{Neill:2018uqw} with the falling $Q^2$ through successive splittings (in an evolution that reminds us of out-of-equilibrium physics with $Q^2$ playing the role of time). Entropy is also generated as the hadrons produced in the final state, often unstable, successively decay to 
lighter, more stable ones~\cite{CarrascoMillan:2018ufj,Llanes-Estrada:2017wip}.

What seems to be missing in the literature is a discussion of the fragmentation functions leading to an identified final-state hadron after the collision at the parton-level has run its course and hadronization occurs, that is, a study of the entropy of Fragmentation Functions. In this work we will carry out a small exploration thereof. 
We will examine the Shannon entropy,
\begin{equation}\label{Shannon}
S= - \sum_i p_i \log p_i 
\end{equation}
in section~\ref{sec:S}, where we report calculations on the known fragmentation functions, that we briefly recall
in subsection~\ref{subsec:FFs}; our estimates provide upper bounds for this $S$, and we try to model how the future
analysis of fragmentation hadrons may change those bounds by educated guessing. 
In section~\ref{sec:FFvspdf} we turn to a comparison of FFs and pdfs, both direct and via the Kullback-Leibler divergence,
 related to the entropy and defined in subsec.~\ref{subsec:cont}.

In brief, the entropy of Eq.~(\ref{Shannon}) is the expectation value of the information in a probability distribution.
Given a set of $I$ objects with two possible states ({\it e.g.} open/closed boxes), or bits of information,
the probability of a configuration scales as  
$p=  \left(\frac{1}{2} \right)^I$, so that the information gained by measuring one of the bits can be quantified 
by the number $I = - \log_2(p_i)$.
The average of this information gauge over the entire distribution, upon changing the base-2 logarithm to base $e$ (with information measured in \emph{nats} instead of bits) is hence Eq.~(\ref{Shannon}), the classical Shannon entropy, whose quantum version, the Von Neumann entropy,  $S=-{\rm Tr} (\rho \log\rho)$, is probably more familiar to subatomic physicists.

\subsection{Fragmentation functions and parton distribution functions} \label{subsec:FFs}

Parton distribution functions are extracted from deeply inelastic scattering (DIS) experiments via the structure functions, for example
\begin{equation} \frac{d\sigma(e^- h \rightarrow e^-X)}{dxdy_{\rm l}}= 
\frac{2\pi \alpha^2}{Q^4}s[1+(1+y_{\rm l})^2 F_2(x)] 
\end{equation}
that are given in terms of the pdfs $f_i$ for the particles and $\bar{f}_i$ for the antiparticles by 
\begin{equation} \label{struct}
F_2{(x)}=\sum_{i}e_i^2x[f_i(x)+\bar{f}_i(x)]\ ,
\end{equation}
with $x$ being the same as $x_i=p_i/P$ the momentum fraction of the hadron carried by the $i^{\rm th}$ parton, and $e_i$ its charge.
An important sum rule that pdfs satisfy is that induced by the addition of each parton's momentum to yield the hadron's
total,
\begin{equation}
\sum_{i} \int xf_i(x) dx = 1\ .
\end{equation}
This sum rule can be interpreted as the normalization of a continuous probability distribution given by $p(i,x):=x f_i(x)$. 
The flavor-sum rules could be chosen instead, but in comparing with fragmentation functions, this momentum sum rule seems more straightforward to us.

Analogously, fragmentation functions~\cite{Osborne:2002dx} can be extracted, for example, from the cross-section for electron-positron 
collisions
\begin{equation} 
\frac{d\sigma(e^-e^+\rightarrow hX)}{dz}= \frac{4\pi\alpha^2}{3Q^2}z^2 \hat{F_2}(z)
\end{equation}
that depend on structure functions analogous to those of Eq.~\ref{struct}, given in terms of the fragmentation functions,
\begin{equation}
\hat{F_2}{(z)}=-3\sum_{q}e_q^2\frac{D^q(z)+D^{\bar{q}}(z)}{z^2}\ .
\end{equation}
Here, $D^q(z)$ represents, in the parton model, the probability that a parton $q$ fragments into the hadron being considered in the final state, $h$, which carries a momentum fraction $z=p_h/p_q$.
These fragmentation functions also satisfy a momentum sum rule,
\begin{equation} \label{momsumrule}
\sum_{h}\int_{0}^{1}zD_h^q(z)dz=1\ . 
\end{equation}
The sum runs over all possible hadrons in which the parton can fragment, and the integral over all possible momentum fractions must yield the original parton momentum, that is, 1 upon normalizing by $p_q$.

\begin{figure*}[h]
\begin{center}
\includegraphics[width=\textwidth]{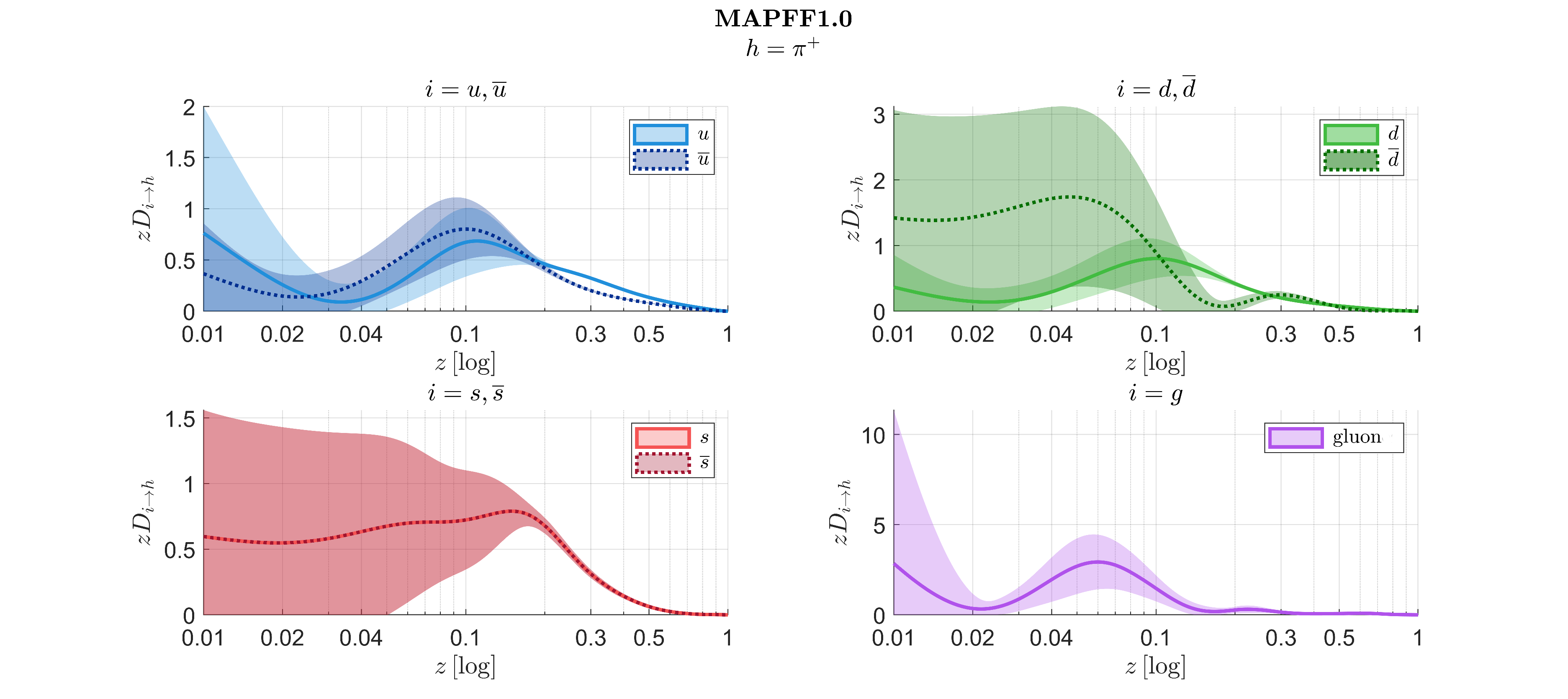}\\
\includegraphics[width=\textwidth]{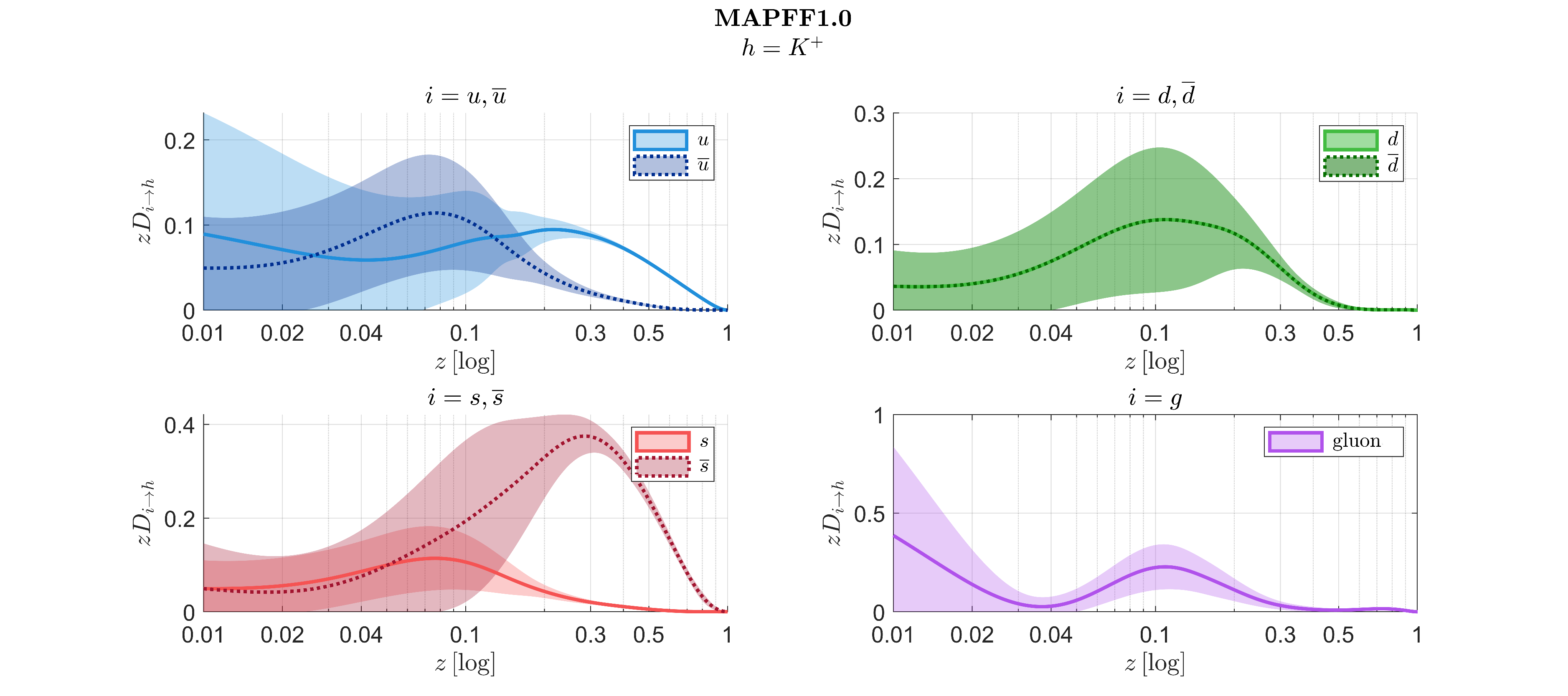}\\
\includegraphics[width=\textwidth]{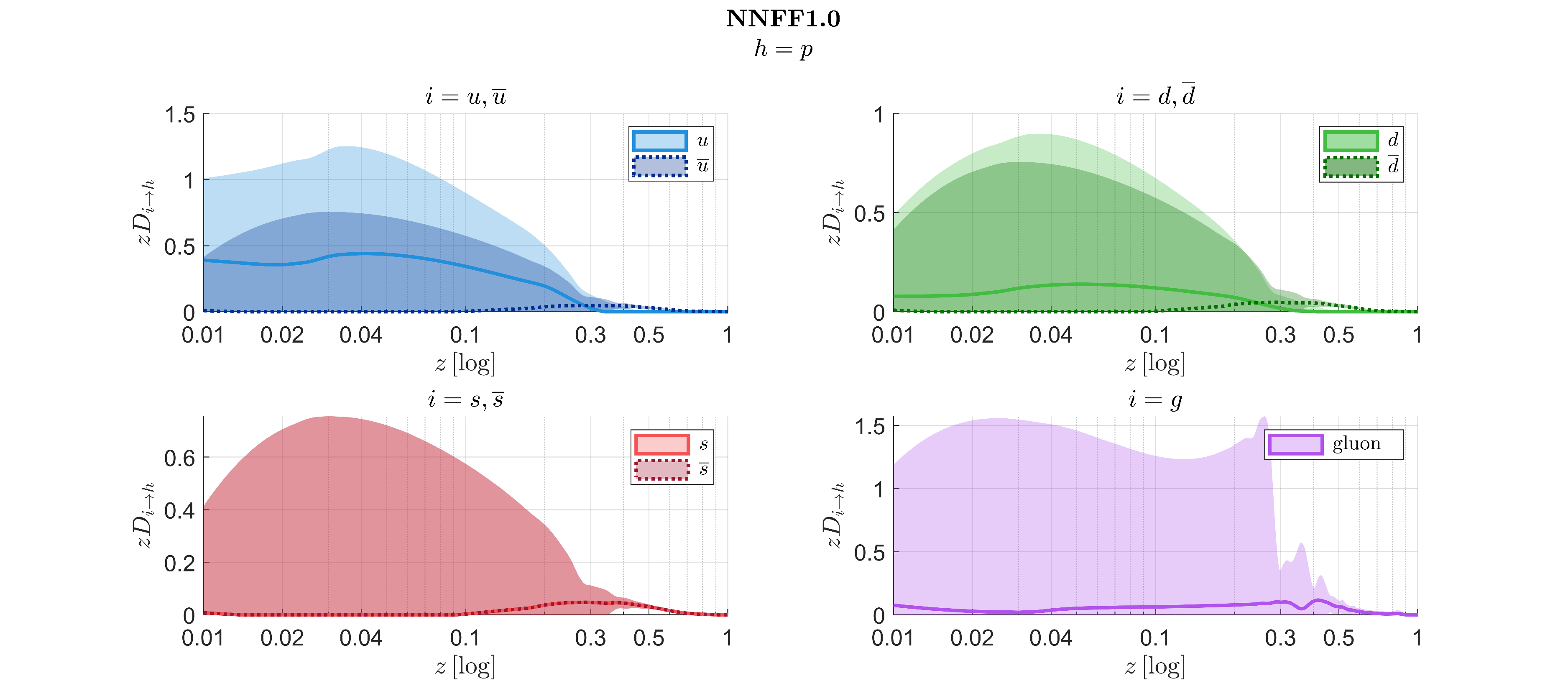}
\end{center}
\caption{\label{fig:FFs}
Fragmentation functions for a pion (top panel), a kaon (middle panel) and a proton (lowest panel) 
fragmenting from the indicated parton, plot from NNPDF and MAPFF public fits~\cite{Bertone:2017tyb,Khalek:2021gxf,AbdulKhalek:2022laj}.
}
\end{figure*}

Parton distribution functions are well known, but for reference we plot in figure~\ref{fig:FFs}
one of the recent determinations of the less widely discussed fragmentation functions, obtained
from neural network fits~\cite{Bertone:2017tyb,Khalek:2021gxf,AbdulKhalek:2022laj} to various experimental data, showing their current large uncertainty bands. Further interesting sets have been reported in the literature~\cite{deFlorian:2014xna,Borsa:2022vvp} but we limit ourselves to relatively low-orders in QCD corrections to avoid 
the Wilson coefficients of the perturbative expansion forcing the nonperturbative distribution function to become negative, 
hiding its nature as a parton probability density, which is asymptotically a correct interpretation, but can be problematic at finite $Q^2$ when QCD corrections are included.

\newpage
\subsection{Relations among pdfs and FFs}

The similarity between pdfs and FFs is due to the basic physics processes that they represent,
$h \to q + X$ and $ q \to h +X$, that are clearly related by crossing.
This did not escape early authors~\cite{Drell:1969wb}, who noted that the two processes, deep inelastic
scattering and fragmentation in $e^-e^+$,
were related by analytic continuation in the kinematic variables.

In current notation, the resulting relation of Drell-Levy-Yan that relates
the fragmentation function $D$ with the parton distribution function $f$ (for the
same hadron and the same parton) reads
\begin{equation}
D(z) = z\ f\left( \frac{1}{z} \right)\ .
\end{equation}
This relation is not particularly useful, since for $z\in [0,1]$,
$\frac{1}{z}\in [1,\infty)$ is unphysical except at $z\to 1$, so that it relates
each of the functions with an analytic extension of the other. (Perhaps this could be tested
on the deuteron or with CLAS data~\cite{Paukkunen:2020rnb}, where $z>1$ might be at hand, but we have not investigated it
as this has not entered the mainstream, where $z\in[0,1]$.)

As discussed by Barone, Drago and Ma~\cite{Barone:2000tx}, Gribov and Lipatov~\cite{Gribov:1972rt}
found a relation between splitting kernels that is usually taken to imply a relation between pdfs
and FFs, the so called   Gribov-Lipatov ``reciprocity'' relation 
\begin{equation}
3 D(z) = f(z)
\end{equation}
but whose theoretical status should rather be considered that of a model.

 Barone, Drago and Ma~\cite{Barone:2000tx} presented then a more suitable expression
\begin{equation} \label{BDMrelation}
D(z)\simeq zf\left( 2-\frac{1}{z}\right)
\end{equation}
that relates the pdf and the FF within their physical domains, that should become exact
in the limit of $z\to 1$, and that apparently should be a good approximation for, say, $z > 0.6$.
Two decades back the scarce quality data did not allow for a good test of this relation, 
and we use this opportunity to employ modern fits of the involved functions to see whether any
progress has been made in empirically establishing this relation.

\newpage
\section{Entropy of parton distribution or fragmentation functions}\label{sec:S}
\subsection{Entropy of continuous distributions} \label{subsec:cont}
A technical problem that arises in directly applying Eq.~(\ref{Shannon}) for the Shannon information entropy to 
a continuous probability distribution is a failure of the simplest discretization. When taking
the continuous probability distribution (here $D(z)=\sum_i D_i(z)$ for a given parton but summed over final-state hadrons)
and mapping it into an approximate, discrete probability set,
$zD(z)\to \left\{ p_1,p_2, \dots p_i, \dots p_N \right\}$; \ \ \ \ \ $\sum_i p_i =1$ 
we find a divergence with the cardinal $N$ of that set (the number of discretization points),
\begin{equation}
S=-\sum_{i=1}^N \left (p_i\log p_i\right)   \underset{N\to \infty}{ \sim}
 \log N \xrightarrow[ N\to \infty]{ } \infty  \ .
\end{equation} 

There is a simple generalization of Eq.~(\ref{Shannon}) to a continuum distribution $F(x)$ (that can represent 
either $xf(x)$ for a pdf, or $zD(z)$ for an FF),
\begin{equation} \label{contS}
S[F]\coloneqq - \int F(x)\ln F(x)\ud x
\end{equation}
but because continuous probability densities can exceed the value $F=1$ in a small-enough interval, 
there can be contributions to this entropy that are negative. This means that there can be occasional
sets that for whatever reason give a negative overall value, which is very counterintuitive for a variable
representing any sort of entropy.

To interpret Eq.~(\ref{contS}), we need to resort to the Kullback-Leibler divergence~\cite{Kullback:1951zyt}.
This divergence takes as argument a second distribution $Q$ and tries to quantify how much does the first, $P$, differ 
from this $Q$, taken as reference:
\begin{equation}
D_\mathrm{KL}(P\Vert Q)\coloneqq\sum_xp(x)\log\dfrac{p(x)}{q(x)}\ .
\end{equation}

It readily admits the continuum generalization
\begin{equation} \label{KLEq}
D_\mathrm{KL}(F\Vert G)=\int f(x)\log\dfrac{f(x)}{g(x)}\ud x
\end{equation}
and then, a typical continuum expression such as Eq.~(\ref{contS})
can be understood as the Kullback-Leibler divergence of the $F(x)$ distribution 
to the uniform one (which is that with least information).

\begin{figure*}[h]
\includegraphics[width=0.95\textwidth]{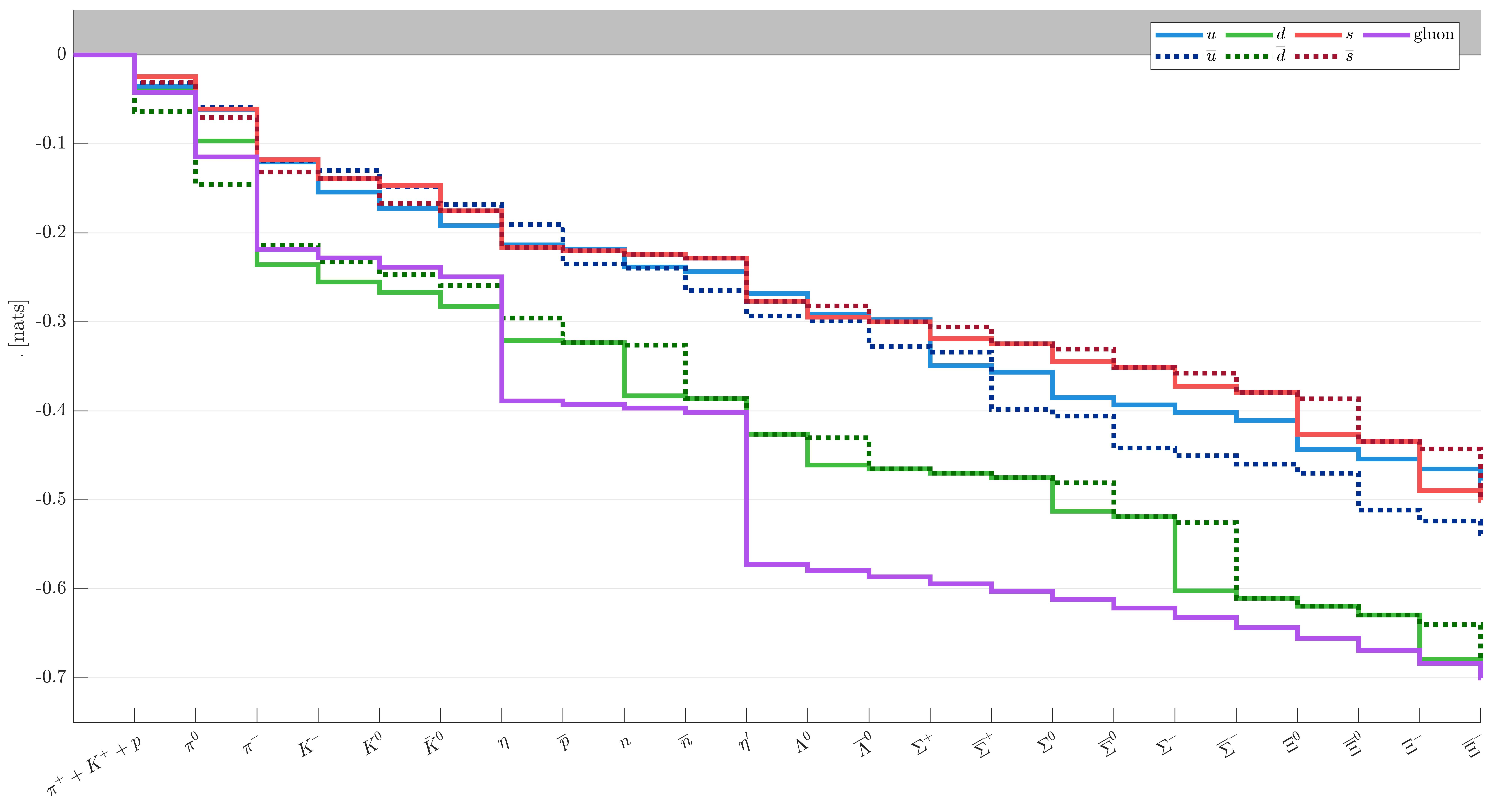}
\caption{\label{fig:escalonada}  Upper bound to the Shannon entropy of fragmentation functions (each line for a different parton, from the NNPDF and MAPFF sets) employing Eq.~(\ref{maxS}) for those hadrons that would not have been measured up to a given point, and the $SU(3)$ quark model relations.}
\end{figure*}

\subsection{Limitation to an upper bound on the entropy when data is insufficient}

To saturate the momentum sum rule in Eq.~(\ref{momsumrule}), the sum needs to be extended 
over \emph{all hadrons} $h$ that can fragment out of the parton $q$ under consideration.

But experimental efforts have been at best identifying a part of the low-lying hadrons (in practice, $\pi^\pm$, $K^\pm$, $p$ and little more), let alone all the unstable 
ones that decay onto these and never reach the detector, so how can an appropriately normalized
distribution function be built with insufficient data?

Our approach~\cite{CarrascoMillan:2018ufj} is to split the fragmentation function into 
a piece that sums over all those hadrons that are specifically identified in a given experiment,
and one that is supposed to represent the unidentified ones, perhaps even unknown, and that 
completes the normalization of the momentum sum rule to be unity,
\begin{equation} \label{fragsplit}
D_q(z) = D_q^{\rm measured}(z) + D_q^{\rm unknown}(z)\ .
\end{equation}

That is, if the momentum sum rule is found to yield $p<1$ in an experiment with only a few
hadrons measured, the rest $1-p$ is assigned to the unknown pieces.

Because we ignore the shape of the unknown fragmentation functions summing the second term of Eq.~(\ref{fragsplit}),
we must renounce to fully compute the entropy from experimental data.
We are forced to assume maximum ignorance of what has not been measured, and assign 
a uniform distribution of $D_q^{\rm unknown}$ over remaining hadrons and momentum fraction $z$, so that
\begin{eqnarray} \label{maxS}
\underbrace{\sum_{h}}_{\rm measured}\int_{0}^{1}zD_h^q(z)dz &=& p<1  \\ 
\underbrace{\sum_{h}}_{\rm unknown}\int_{0}^{1} zD_h^q(z)dz &:= & 1-p \ .
\end{eqnarray}

That means that the entropy computed with this distribution function is an upper bound on the actual
entropy $S$: the uniformity assumption for the remainder is equivalent to a maximization of the possible entropy.
Of course, if successive data taking adds knowledge of $D_h^q$ for some hitherto unmeasured hadron,
that upper bound on $S$ improves.

We have carried out a simulation of what that future improvement might be, and report it in figure~\ref{fig:escalonada}.

\begin{landscape}
\begin{table}[]
\centering
\resizebox{1.3\textwidth}{!}{%
\begin{tabular}{@{}ccccccccc@{}}
\toprule
\textbf{$\boldsymbol{h}$} &
  \textbf{Quark model wavefunction} &
  \textbf{$\boldsymbol{D_{u\to h}}$} &
  \textbf{$\boldsymbol{D_{\overline{u}\to h}}$} &
  \textbf{$\boldsymbol{D_{d\to h}}$} &
  \textbf{$\boldsymbol{D_{\overline{d}\to h}}$} &
  \textbf{$\boldsymbol{D_{s\to h}}$} &
  \textbf{$\boldsymbol{D_{\overline{s}\to h}}$} &
  \textbf{$\boldsymbol{D_{g\to h}}$} \\ \midrule
$\pi^0$ &
  $\tfrac{1}{\sqrt{2}}(u\overline{u}+d\overline{d})$ &
  {\color[HTML]{9A0000} $\tfrac{1}{\sqrt{2}}D_{u\to\pi^+}$} &
  {\color[HTML]{9A0000} $\tfrac{1}{\sqrt{2}}D_{u\to\pi^+}$} &
  {\color[HTML]{9A0000} $\tfrac{1}{\sqrt{2}}D_{\overline{d}\to\pi^+}$} &
  {\color[HTML]{9A0000} $\tfrac{1}{\sqrt{2}}D_{\overline{d}\to\pi^+}$} &
  {\color[HTML]{3166FF} $D_{s\to\pi^+}$} &
  {\color[HTML]{3166FF} $D_{s\to\pi^+}$} &
  {\color[HTML]{9B9B9B} $D_{g\to\pi^+}$} \\
$\pi^+$ &
  $u\overline{d}$ &
  \cellcolor[HTML]{F8BBD0}$D_{u\to\pi^+}$ &
  \cellcolor[HTML]{F9D7AF}$D_{d\to\pi^+}$ &
  \cellcolor[HTML]{F9D7AF}$D_{d\to\pi^+}$ &
  \cellcolor[HTML]{F8BBD0}$D_{\overline{d}\to\pi^+}$ &
  \cellcolor[HTML]{B2EBF2}$D_{s\to\pi^+}$ &
  \cellcolor[HTML]{B2EBF2}$D_{s\to\pi^+}$ &
  \cellcolor[HTML]{CFD8DC}$D_{g\to\pi^+}$ \\
$K^+$ &
  $u\overline{s}$ &
  \cellcolor[HTML]{F8BBD0}$D_{u\to K^+}$ &
  \cellcolor[HTML]{F9D7AF}$D_{s\to K^+}$ &
  \cellcolor[HTML]{B2EBF2}$D_{d\to K^+}$ &
  \cellcolor[HTML]{B2EBF2}$D_{d\to K^+}$ &
  \cellcolor[HTML]{F9D7AF}$D_{s\to K^+}$ &
  \cellcolor[HTML]{F8BBD0}$D_{\overline{s}\to K^+}$ &
  \cellcolor[HTML]{CFD8DC}$D_{g\to K^+}$ \\
$K^0$ &
  $d\overline{s}$ &
  {\color[HTML]{3166FF} $D_{d\to K^+}$} &
  {\color[HTML]{3166FF} $D_{d\to K^+}$} &
  {\color[HTML]{9A0000} $D_{u\to K^+}$} &
  {\color[HTML]{CB841A} $D_{s\to K^+}$} &
  {\color[HTML]{CB841A} $D_{s\to K^+}$} &
  {\color[HTML]{9A0000} $D_{\overline{s}\to K^+}$} &
  {\color[HTML]{9B9B9B} $D_{g\to K^+}$} \\
$\eta$ &
  \begin{tabular}[c]{@{}c@{}}$c_1(u\overline{u}+d\overline{d})$\\      $\qquad +c_2s\overline{s}$\end{tabular} &
  {\color[HTML]{9A0000} \begin{tabular}[c]{@{}c@{}}$\tfrac{c_1}{\sqrt{2}}D_{u\to\pi^+}$\\  $ \qquad   +c_2D_{d\to K^+}$\end{tabular}} &
  {\color[HTML]{9A0000} \begin{tabular}[c]{@{}c@{}}$\tfrac{c_1}{\sqrt{2}}D_{u\to\pi^+}$\\ $\qquad     +c_2D_{d\to K^+}$\end{tabular}} &
  {\color[HTML]{9A0000} \begin{tabular}[c]{@{}c@{}}$\tfrac{c_1}{\sqrt{2}}D_{\overline{d}\to\pi^+}$\\ $\qquad     +c_2D_{d\to K^+}$\end{tabular}} &
  {\color[HTML]{9A0000} \begin{tabular}[c]{@{}c@{}}$\tfrac{c_1}{\sqrt{2}}D_{\overline{d}\to\pi^+}$\\      $\qquad+c_2D_{d\to K^+}$\end{tabular}} &
  {\color[HTML]{9A0000} \begin{tabular}[c]{@{}c@{}}$c_1D_{s\to\pi^+}$\\      $\qquad+c_2D_{\overline{s}\to K^+}$\end{tabular}} &
  {\color[HTML]{9A0000} \begin{tabular}[c]{@{}c@{}}$c_1D_{s\to\pi^+}$\\      $\qquad+c_2D_{\overline{s}\to K^+}$\end{tabular}} &
  {\color[HTML]{9B9B9B} $D_{g\to\pi^+}$} \\
$\eta'$ &
  \begin{tabular}[c]{@{}c@{}}$c_1'(u\overline{u}+d\overline{d})$\\   $\qquad   +c_2's\overline{s}$\end{tabular} &
  {\color[HTML]{9A0000} \begin{tabular}[c]{@{}c@{}}$\tfrac{c_1'}{\sqrt{2}}D_{u\to\pi^+}$\\ $\qquad     +c_2'D_{d\to K^+}$\end{tabular}} &
  {\color[HTML]{9A0000} \begin{tabular}[c]{@{}c@{}}$\tfrac{c_1'}{\sqrt{2}}D_{u\to\pi^+}$\\ $\qquad    +c_2'D_{d\to K^+}$\end{tabular}} &
  {\color[HTML]{9A0000} \begin{tabular}[c]{@{}c@{}}$\tfrac{c_1'}{\sqrt{2}}D_{\overline{d}\to\pi^+}$\\ $\qquad     +c_2'D_{d\to K^+}$\end{tabular}} &
  {\color[HTML]{9A0000} \begin{tabular}[c]{@{}c@{}}$\tfrac{c_1'}{\sqrt{2}}D_{\overline{d}\to\pi^+}$\\      $\qquad+c_2'D_{d\to K^+}$\end{tabular}} &
  {\color[HTML]{9A0000} \begin{tabular}[c]{@{}c@{}}$c_1'D_{s\to\pi^+}$\\      $\qquad+c_2'D_{\overline{s}\to K^+}$\end{tabular}} &
  {\color[HTML]{9A0000} \begin{tabular}[c]{@{}c@{}}$c_1'D_{s\to\pi^+}$\\      $\qquad+c_2'D_{\overline{s}\to K^+}$\end{tabular}} &
  {\color[HTML]{9B9B9B} $D_{g\to\pi^+}$} \\ \midrule
$p$ &
  $uud$ &
  \cellcolor[HTML]{F8BBD0}{\color[HTML]{000000} $2D_{\mathrm{val}\to p}$} &
  \cellcolor[HTML]{B2EBF2}{\color[HTML]{000000} $D_{\mathrm{sea}\to p}$} &
  \cellcolor[HTML]{F8BBD0}{\color[HTML]{000000} $D_{\mathrm{val}\to p}$} &
  \cellcolor[HTML]{B2EBF2}{\color[HTML]{000000} $D_{\mathrm{sea}\to p}$} &
  \cellcolor[HTML]{B2EBF2}{\color[HTML]{000000} $D_{\mathrm{sea}\to p}$} &
  \cellcolor[HTML]{B2EBF2}{\color[HTML]{000000} $D_{\mathrm{sea}\to p}$} &
  \cellcolor[HTML]{CFD8DC}{\color[HTML]{000000} $D_{g\to p}$} \\
$n$ &
  $udd$ &
  {\color[HTML]{9A0000} $D_{\mathrm{val}\to p}$} &
  {\color[HTML]{3166FF} $D_{\mathrm{sea}\to p}$} &
  {\color[HTML]{9A0000} $2D_{\mathrm{val}\to p}$} &
  {\color[HTML]{3166FF} $D_{\mathrm{sea}\to p}$} &
  {\color[HTML]{3166FF} $D_{\mathrm{sea}\to p}$} &
  {\color[HTML]{3166FF} $D_{\mathrm{sea}\to p}$} &
  {\color[HTML]{9B9B9B} $D_{g\to p}$} \\
$\Lambda^0$ &
  $uds$ &
  {\color[HTML]{9A0000} $D_{\mathrm{val}\to p}$} &
  {\color[HTML]{3166FF} $D_{\mathrm{sea}\to p}$} &
  {\color[HTML]{9A0000} $D_{\mathrm{val}\to p}$} &
  {\color[HTML]{3166FF} $D_{\mathrm{sea}\to p}$} &
  {\color[HTML]{9A0000} $D_{\mathrm{val}\to p}$} &
  {\color[HTML]{3166FF} $D_{\mathrm{sea}\to p}$} &
  {\color[HTML]{9B9B9B} $D_{g\to p}$} \\
$\Sigma^+$ &
  $uus$ &
  {\color[HTML]{9A0000} $2D_{\mathrm{val}\to p}$} &
  {\color[HTML]{3166FF} $D_{\mathrm{sea}\to p}$} &
  {\color[HTML]{3166FF} $D_{\mathrm{sea}\to p}$} &
  {\color[HTML]{3166FF} $D_{\mathrm{sea}\to p}$} &
  {\color[HTML]{9A0000} $D_{\mathrm{val}\to p}$} &
  {\color[HTML]{3166FF} $D_{\mathrm{sea}\to p}$} &
  {\color[HTML]{9B9B9B} $D_{g\to p}$} \\
$\Sigma^0$ &
  $uds$ &
  {\color[HTML]{9A0000} $D_{\mathrm{val}\to p}$} &
  {\color[HTML]{3166FF} $D_{\mathrm{sea}\to p}$} &
  {\color[HTML]{9A0000} $D_{\mathrm{val}\to p}$} &
  {\color[HTML]{3166FF} $D_{\mathrm{sea}\to p}$} &
  {\color[HTML]{9A0000} $D_{\mathrm{val}\to p}$} &
  {\color[HTML]{3166FF} $D_{\mathrm{sea}\to p}$} &
  {\color[HTML]{9B9B9B} $D_{g\to p}$} \\
$\Sigma^-$ &
  $dds$ &
  {\color[HTML]{3166FF} $D_{\mathrm{sea}\to p}$} &
  {\color[HTML]{3166FF} $D_{\mathrm{sea}\to p}$} &
  {\color[HTML]{9A0000} $2D_{\mathrm{val}\to p}$} &
  {\color[HTML]{3166FF} $D_{\mathrm{sea}\to p}$} &
  {\color[HTML]{9A0000} $D_{\mathrm{val}\to p}$} &
  {\color[HTML]{3166FF} $D_{\mathrm{sea}\to p}$} &
  {\color[HTML]{9B9B9B} $D_{g\to p}$} \\
$\Xi^0$ &
  $uss$ &
  {\color[HTML]{9A0000} $D_{\mathrm{val}\to p}$} &
  {\color[HTML]{3166FF} $D_{\mathrm{sea}\to p}$} &
  {\color[HTML]{3166FF} $D_{\mathrm{sea}\to p}$} &
  {\color[HTML]{3166FF} $D_{\mathrm{sea}\to p}$} &
  {\color[HTML]{9A0000} $2D_{\mathrm{val}\to p}$} &
  {\color[HTML]{3166FF} $D_{\mathrm{sea}\to p}$} &
  {\color[HTML]{9B9B9B} $D_{g\to p}$} \\
$\Xi^-$ &
  $dss$ &
  {\color[HTML]{3166FF} $D_{\mathrm{sea}\to p}$} &
  {\color[HTML]{3166FF} $D_{\mathrm{sea}\to p}$} &
  {\color[HTML]{9A0000} $D_{\mathrm{val}\to p}$} &
  {\color[HTML]{3166FF} $D_{\mathrm{sea}\to p}$} &
  {\color[HTML]{9A0000} $2D_{\mathrm{val}\to p}$} &
  {\color[HTML]{3166FF} $D_{\mathrm{sea}\to p}$} &
  {\color[HTML]{9B9B9B} $D_{g\to p}$} \\ \bottomrule
\end{tabular}%
}
\caption{Relations used to recreate the fragmentation functions into light mesons and baryons, starting from a few
ones that are already measured, those of $\pi^+$, $K^+$ and $p$. Those with a shaded background are the input ones,
whereas the majority without a background are related to those as an educated guess to compute the previsible improvement
to the upper bound on the entropy.
Color online: we employ red for valence quark combinations, blue for sea quarks, orange for partons with additional suppression, and grey is reserved for gluons.
We omit the $\pi^-,K^-,\overline{K}^0$ antiparticles and all antibaryons, that are immediately obtained from
charge conjugation such as $D_{q\to h}=D_{\overline{q}\to\overline{h}}$.}
\label{tab:baryon}
\end{table}
\end{landscape}

\begin{figure*}[h!]
\includegraphics[width=0.8\textwidth]{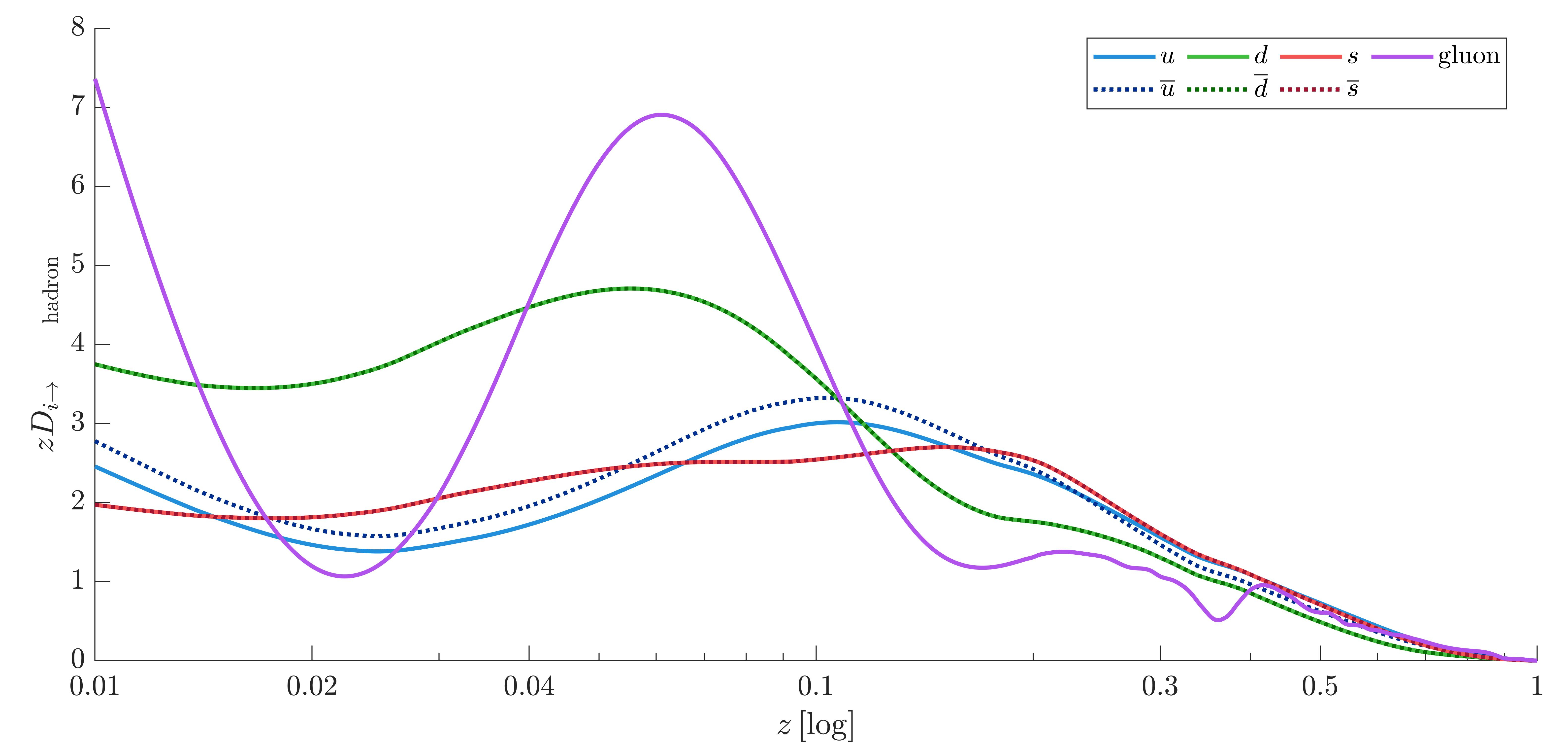}
\caption{  \label{fig:reconstructedD}  Reconstructed aggregated fragmentation functions summing over all low-lying hadrons 
following the rules of table~\ref{tab:baryon}. This is an educated guess of what future experimental progress could bring; upon summing over hadron species, the remaining information is of course only about a momentum distribution, and the bound over the entropy in figure~\ref{fig:escalonada} should hopefully be converging to the actual entropy of that distribution.}
\end{figure*}

The contribution to the entropy of Eq.~(\ref{contS}) due to the identified charged hadrons, with NNPDF fits for $D_q^h$,
is shown as the first step to the left of the figure. Each possible fragmenting parton is distinguished by a different type of line as specified in the caption. For this first step, Eq.~(\ref{maxS}) has been used.

The figure is then organized so that hadrons from the lowest-mass baryon octet and meson nonet (both $\eta$ and $\eta'$, mixed with  $\theta(\eta,\eta')= -15.5^{\rm o}$, are included~\cite{Bramon:1997va}) appear towards the right, broadly ordered by increasing mass.

In simulating how the upper bound on the entropy might descend with future knowledge, we have coarsely represented
each successive function $D_h^q$ by a known fragmentation function, assigning them employing
$SU(3)$ relations linking the various flavors of hadrons and quarks, and employing the customary
valence/sea separation. For example, the as of yet unmeasured fragmentation of a down quark to 
a neutral Kaon $D_{d\to K^0} $, is set to be equal to that of an up quark fragmenting to a positively charged kaon  $D_{u\to K^+}$  (already extracted). An extensive list of how these relations were applied can be found in table~\ref{tab:baryon}.

In view of Fig.~\ref{fig:escalonada}, we therefore propose the upper bound on the entropy $S$ as
a salient number that quantifies progress on each of the fragmentation functions. 

For completeness,  we plot in Fig.~\ref{fig:reconstructedD} 
the ``reconstructed'' aggregate fragmentation function that sums over the hadron species just discussed~\footnote{How each of them contributes to that distribution can be seen in the sequence of graphs reported to the XVth International Conference on Quark Confinement and the Hadron Spectrum, \href{https://indico.uis.no/event/2/contributions/576/attachments/210/313/myimage.gif}{downloadable at its website}.}.
That aggregated reconstruction eliminates (sums over) hadron flavor information to display only the distribution of 
the hadron momentum fraction taken from the fragmenting parton.


\section{Comparison between $D$s and $FF$s} \label{sec:FFvspdf}
\subsection{Graphical examination}
Next, we turn to a cursory examination of Eq.~(\ref{BDMrelation}),  as well as that of its further
simplification as $z\to 1$ 
\begin{equation}\label{BDMrelationapprox}
D(z)\simeq zf(z)
\end{equation} 
that comes about because
\begin{equation}
z= \frac{1}{1-(1-1/z)} \simeq 2-\frac{1}{z} 
\end{equation}
which holds up to two orders of the Taylor series.
 
\begin{figure*}
\begin{center}
        \includegraphics[width=0.85\columnwidth]{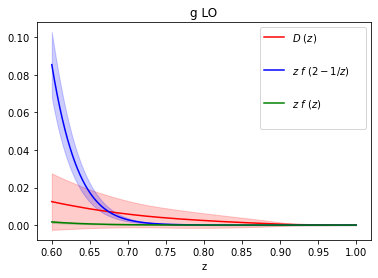} 
        \includegraphics[width=0.85\columnwidth]{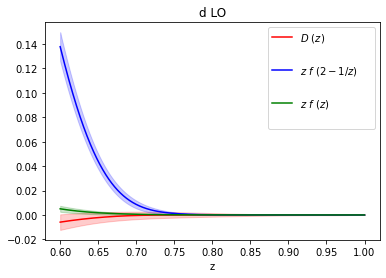}\\
        \includegraphics[width=0.85\columnwidth]{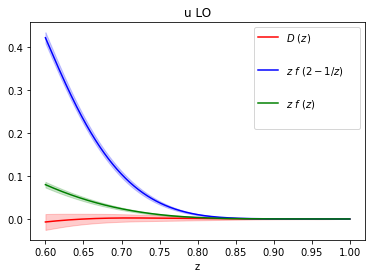} 
        \includegraphics[width=0.85\columnwidth]{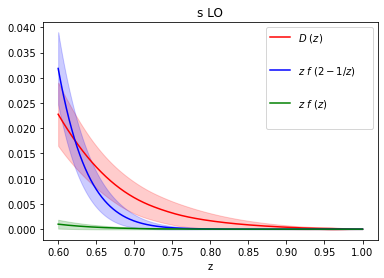}
\caption{\label{fig:compareRelations1} Visualization of Eq.~(\ref{BDMrelation}) and~(\ref{BDMrelationapprox}) for the proton.}
\end{center}
\end{figure*}

In the first place, we plot all three functions for DIS on the proton and proton fragmentation
(using the NNPDF parametrizations, with scale 
$Q^2=100 \rm{GeV^2}$) in figure~\ref{fig:compareRelations1}.

\begin{figure*}
\begin{center}
        \includegraphics[width=0.85\columnwidth]{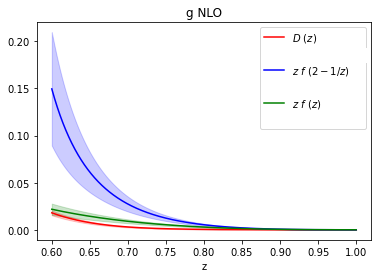} 
        \includegraphics[width=0.85\columnwidth]{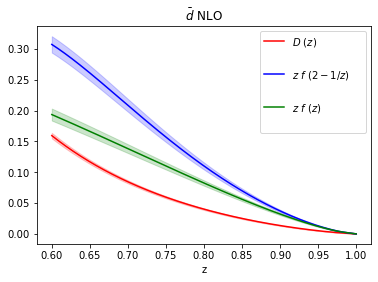}\\
        \includegraphics[width=0.85\columnwidth]{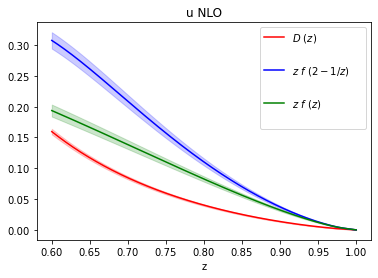} 
        \includegraphics[width=0.85\columnwidth]{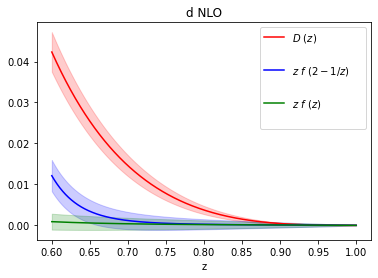}
\caption{\label{fig:compareRelations2}Visualization of Eq.~(\ref{BDMrelation}) and~(\ref{BDMrelationapprox}) for  $\pi^+/\pi^-$.}
\end{center}
\end{figure*}

Then, in figure~\ref{fig:compareRelations2}, we present the equivalent plots for the
charged pion. 
Here, we have used the fit to $\pi^+$ data, with scale $Q^2=100 \rm{GeV^2}$, obtained by the Jefferson Lab Angular Momentum
collaboration~\cite{Moffat:2021dji} that has simultaneously fit parton and fragmentation densities,
employing traditional template fits with power-law end-point behaviour $\alpha (1-z)^\beta$. 

In both cases we do not perceive by eyeball that either of the relations in Eq.~(\ref{BDMrelation})
or Eq.~(\ref{BDMrelationapprox}) are satisfied better than the fact that all three bands smoothly converge to zero as 
$z\to 1$.

\begin{figure*} 
\begin{center}
        \includegraphics[width=0.85\columnwidth]{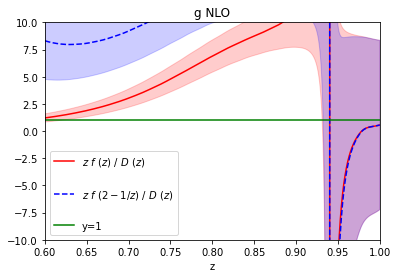} 
        \includegraphics[width=0.85\columnwidth]{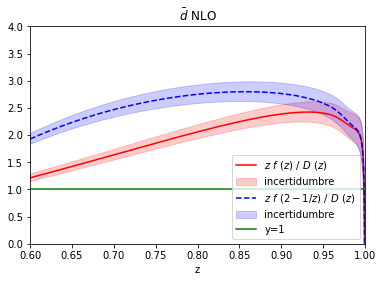}\\
        \includegraphics[width=0.85\columnwidth]{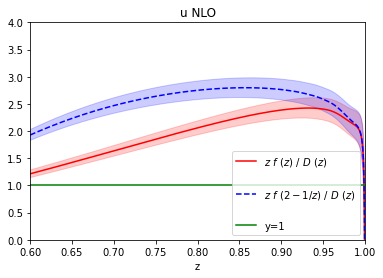} 
        \includegraphics[width=0.85\columnwidth]{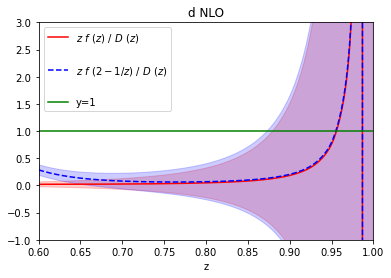}
\caption{Ratios of the adequate evaluation (Eqs.~(\ref{BDMrelation}) and~(\ref{BDMrelationapprox}) ) of the parton distribution functions and the fragmentation functions do not seem to be very near unity in the region near the endpoint $z\to 1$.\label{ratios}}
\end{center}
\end{figure*}

This is corroborated by taking ratios between $D$s and pdfs (plot in figure~\ref{ratios}).  
There is no definite evidence that the ratios are near unity: two of the plots have huge errors
and although compatible with unity, they are inconclusive; and two are clearly incompatible, 
even as a limit as $z\to 1$, where the ratio seems to overshoot.

\newpage

\subsection{Quantification by KL divergence}
To quantify this perceived separation between the fragmentation function and the corresponding small modification of the parton distribution function, we make use of the Kullback-Leibler divergence of Eq.~(\ref{KLEq}). 
To get a feeling for its typical value, we first carry out an exercise with two simple pairs of normal probability distributions (unrelated to physical FFs) with different parameters. 

\begin{figure*}
\begin{center}
\includegraphics[width=0.9\columnwidth]{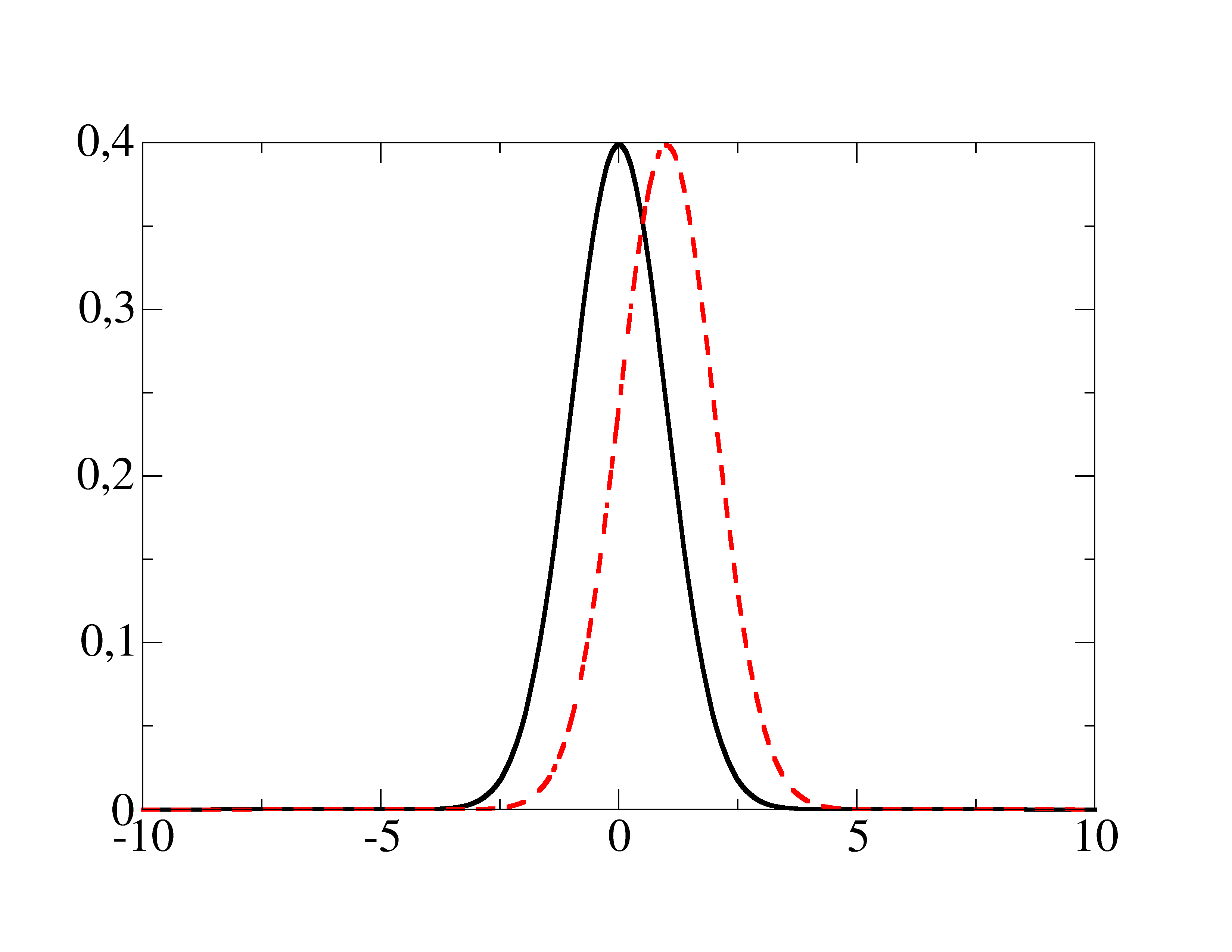}
\includegraphics[width=0.9\columnwidth]{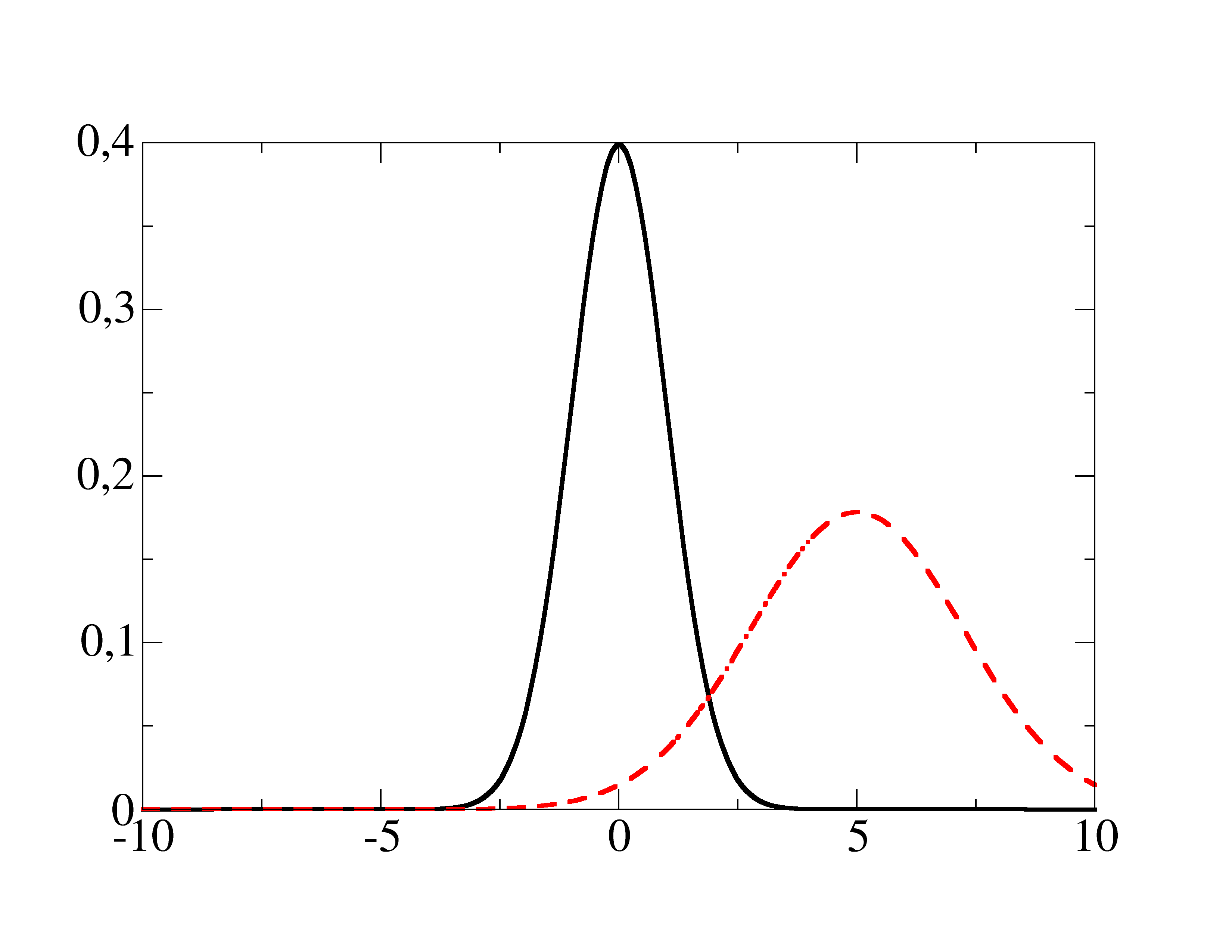}
\caption{\label{toydist} Two pairs of normalized probability distributions, similar and dissimilar, as 
given in Eq.~(\ref{similardists}) and (\ref{dissimilardist}), which have respective Kullback-Leibler divergences
of 0.12 and 1.63.
}
\end{center}
\end{figure*}

First, the Kullback-Leibler divergence among
\begin{equation} \label{similardists}
f(x) = \frac{1}{\sqrt{2\pi}} e^{-\frac{1}{2}x^2}\ \ \ g_1(x) = \frac{1}{\sqrt{2\pi}} e^{-\frac{1}{2}(x-1)^2}
\end{equation}
which are two distributions that overlap at 1$\sigma$, and thus relatively likely, turns out to be
$D_{\rm KL}(f,g_1) = 0.12 $.
On the other hand, if we substitute $g_1$ by 
\begin{equation}\label{dissimilardist}
g_2(x) =  \frac{1}{5\sqrt{2\pi}} e^{-\frac{1}{2}(x-5)^2}\ ,
\end{equation}
that is significantly different from $f$ (both pairs of distributions 
are plot in figure~(\ref{toydist})), 
we have $D_{\rm KL}(f,g_2)= 1.63$; thus, we expect values of $D_{KL}$ near 1 or above to signify that the distributions
are not alike, while values well below 1 mean that the distributions are more similar.

Having a feeling for the divergence, we turn to the physical relation among fragmentation functions
and parton distribution functions, with the result given in table~\ref{tab:KL}.

\begin{table} 
\caption{Kullback-Leibler divergence for two physical relations among pdfs and FFs for the charged pion from Jefferson Lab Angular Momentum collaboration fits, considering only the integration interval $z\in(0.7,1)$ where the relations have been proposed to be valid. We find them to  not be well satisfied. \label{tab:KL}}
\begin{tabular}{||c c c||} 
    \hline
    Parton & $D_{KL}(zf_\pi(z)||D^\pi(z))$ & $D_{KL}(zf_\pi(2-1/z)||D^\pi(z))$      \\ [0.5ex] 
    \hline\hline
    gluon & 2,28 & 7,70 \\ 
    \hline
    up & 1,19  &  2,36\\
    \hline
    anti-down & 1,19 & 2,36  \\
    \hline
    down & 6,64 &  5,28  \\
    \hline
    \end{tabular}
\end{table}

We see there that the Kullback-Leibler divergence gives values of order 1 or above for all considered partons 
(we are using the Jefferson Lab Angular Momentum (JAM) evaluations for the charged pion) so that none of these relations
looks very promising at the present time. 

\newpage
\subsection{Similar power laws}
\begin{figure}
\includegraphics[width=\columnwidth]{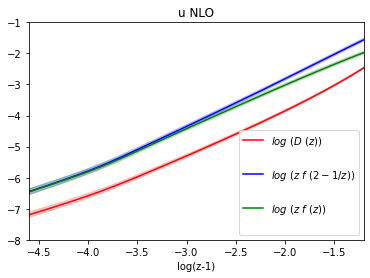}\\
\includegraphics[width=\columnwidth]{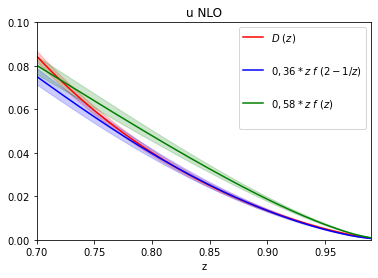}
\caption{\label{fig:power}
Top plot: FFs vs pdfs in a log-log plot coming into the end-point $z\to 1$ (notice the this makes the $OX$ axis opposite 
to its usual representation, so that the functions fall towards the left). This suggests that the power-law is the same
for both types of functions, and that the reason that the Kullback-Leibler divergence in table~\ref{tab:KL} perceives a difference is due to a multiplicative constant.
Bottom plot: FFs vs pdfs multiplied by a fixed constant, as suggested by the top plot, now in the usual scale,
showing great ressemblance.
}
\end{figure}

However, we do find, always looking at the JAM distributions~\cite{Moffat:2021dji}, that the fragmentation functions
do have a similarity to the parton distribution functions. This is observed in figure~\ref{fig:power}.
In it, we once more turn to the $\pi^+$ fragmentation functions as well as parton distribution functions from the 
NLO parametrizations of the JAM collaboration~\cite{Moffat:2021dji} at a scale $Q^2=100 \rm{GeV^2}$. 
We look, in particular, at the $u$-quark (which is a valence parton for the positively charged pion), so that we will be examining relations among $f_\pi^u(x)$ and $D_u^\pi(z)$.

The top plot is a logarithmic plot of the three distribution functions discussed by Barone, Drago and Ma~\cite{Barone:2000tx}
against $\log(z-1)$, so that the functions run down towards the left of the plot. In such a log-log plot,
a power-law appears as a straight line, and this is indeed the case. 
Moreover, all three are rather parallel lines, and that suggests that the failure of the Barone-Drago-Ma relations at least in certain fits of data can be due to overall multiplicative constants in the region $z\simeq 1$, since the distributions
with the appropriate arguments seem to be coming into the end point with the same slope.

Both the original $zf(2-1/z)$ as well as its further approximation $zf(z)$ seem to track the fall-off of the fragmentation function as proposed by those authors.
The slopes of the three quasi-straight lines shown (the power law exponent of $\alpha_i (z-1)^{\beta_i}$)
are, respectively, 1.48 for $D(z)$, againsst   1.52 for      $zf(2-\frac{1}{z})$, which are rather close,
and even  a not too far 1.3  for $zf(z)$. 

Thus, a multiplicative constant seems to be the difference between these distributions. This can be appreciated 
in plain sight in the lower plot of figure~\ref{fig:power}, where an {\it ad-hoc} constant multiplies the  FF and puts the the three lines on top of each other. The Kullback-Leibler divergence between pairs of functions sees a significant drop, as expected. 

We presently ignore what the theoretical reason for this may be, but we hope it can be studied in the future by the 
collaborations addressing fragmentation functions and parton distribution functions in a unified manner.

\section{Conclusions and outlook} 

In this contribution we have deployed two very well-known estimators of information theory, 
the Shannon information entropy and the Kullback-Leibler divergence, to assess fragmentation
functions. 

We have found that the entropy is one adequate number to quantify future progress on knowledge 
of the fragmentation functions, binding $S$ from above with a uniform, maximum-entropy distribution
for the unknown channels of fragmentation, as shown in figure~\ref{fig:escalonada}.

One issue that we have not discussed in length but that may play a role in future analysis is the fact that
fragmenting partons may yield excited hadron resonances that are now very well known to exist~\cite{Workman:2022ynf}
but were not contemplated in early quark analysis of the sixties. These excited hadron resonances (such as 
$N^*,\Delta^*\to p\dots$)
undergo one or more secondary decays to Gell-Mann's quark-model basic hadrons, and add to their fragmentation function~\cite{Adamov:2000is} (for example, here that of the proton),
even when the time-scales involved may be different and it would be more convenient to distinguish them as
separate fragmentation functions.

We have also examined the relations among parton distribution and fragmentation functions proposed in the literature
for the same parton and hadron, and found them to not be well satisfied with current sets.

We have found at least in one case, that of $D_{u\to \pi^+}$, that the reason is not a power-law mismatch 
between the parton distribution and the fragmentation function, but rather a multiplicative constant (an upward
shift in a logarithmic plot). While data sets tend to have larger errors near the $z=1$ endpoint, 
we believe that with modest effort, a future systematic analysis of this observation can be carried out as
those uncertainty bands come somewhat down, to see whether it stands further tests.
Measurement of the power-laws of fragmentation functions has long been known to offer a window into 
hadron dynamics~\cite{Brodsky:1977bu} and we hope to see future progress.

\section*{Acknowledgments}
We thank Dr. Pia Zurita for useful comments.\\
This project has received funding from the European Union's Horizon 2020 research and innovation programme under grant agreement No 824093; grants  MICINN: PID2019-108655GB-I00, PID2019-106080GB-C21 (Spain); UCM research group 910309 and the IPARCOS institute.


\printcredits



\end{document}